# Teaching STEM Courses using Ignatian Pedagogy

Gintaras Dūda
Professor of Physics
Creighton University

**Abstract**

Despite the impact of the Jesuit educational endeavor on the rise of science, the Ignatian Pedagogical Paradigm (IPP), the signature Jesuit pedagogy, is not frequently used to teach courses in science, technology, engineering and mathematics (STEM), and very little literature exists documenting any such attempts. In this paper, I describe a framework for how to apply the IPP to STEM courses using active-engagement strategies and assessment tools from disciplinary educational research (DBER). I provide three examples of how I have implemented the IPP in physics courses at various levels in the curriculum at a Jesuit University complete with assessment results that demonstrate student learning. I stress that beyond the technical, cyclical elements of the IPP, a truly Ignatian course needs to pay close attention to Jesuit charisms such as cura personalis, magis, and educating men and women with and for others.

**Introduction**

Natural science and mathematics have been an important component of Jesuit education as early as initial drafts of the Ratio Studiorum in 1586. Christopher Clavius S.J., chief proponent of the Gregorian calendar, is often cited as the architect of this integration of mathematics and the mathematical sciences into the more standard humanistic education of the day.[1] Superior General Peter-Hans Kolvenbach, S.J., in an address to Georgetown trustees, quotes Clavius saying, "the mathematical disciplines in fact require truth, delight in truth, and honour truth ... there can be no doubt that they must be conceded the first place among all the other sciences."[2] This integration had profound consequences down the centuries, as evidenced by Jesuit scientists such as Giovanni Battisti Riccioli, Christoph Scheiner, Matteo Ricci, and Francesco Maria Grimaldi, all of whom left lasting impacts on the science of the 16th and 17th centuries. As historian of science George Sarton observed in 1940, "One cannot talk about mathematics in the 16th and 17th centuries without seeing a Jesuit at every corner."[3]

In fact, some scholars have argued that Jesuit education helped foster the rise of "science" as the discipline we know it as today. Historian Rivka Feldhay argues "that the Jesuit educational system - whose origins are traceable to the first years of the Society in the 1540s - has allowed for, and even to a certain extent encouraged, the differentiation of the mathematical disciplines and certain parts of natural philosophy from other fields of learning, and their emergence as a specific area of studies and research similar enough to what is recognized by us as 'science.'".[4] Mordechai Feingold further supports this idea by proposing that Jesuits made a key contribution to "the two poles of modern science: the mathematization of natural philosophy and experimental science."[5]

In more modern times the nature and distinctiveness of Jesuit education was spelled out in a document in 1986 crafted after four years of intensive meetings and collaboration by the International Commission on the Apostolate of Jesuit Education (ICAJE).[6] This document acknowledges that in modern times, mathematics and the sciences are increasingly seen as even more crucial elements of higher education. The report admits that "the course of studies has been altered by modern advances in science and technology"[7], though it laments that this has occurred somewhat at the expense of the humanistic disciplines. In 1993, a more succinct document, "Ignatian Pedagogy: A Practical Approach"[8] was released which sparked a new interest in the signature Jesuit pedagogy which endures today. Yet despite this resurgence in interest in and the renewed importance of the Ignatian Pedagogical Paradigm (IPP) at Jesuit universities, the use of the IPP is still extremely limited today in science and mathematics courses even at Jesuit universities or at the very least exists in isolated pockets with no broader communication and sharing. This is ironic particularly given the important role that Jesuits and Jesuit higher education played in the rise of science.

One motivation to write this article was that I found that the literature exploring the use of the IPP within science and mathematics courses, or more broadly within STEM disciplines, at the university-level is almost non-existent. A few pieces here and there exist in very specific contexts, but no grand synthesis of principles of teaching science using the IPP exists. For example, a scholar in the Philippines explored teaching science communication and risk management using Ignatian Pedagogy, but not teaching science itself.[9] Work at Loyola Marymount University examined using Ignatian Pedagogical principles to help faculty transition STEM courses online.[10] Research at Creighton University focused on applying Ignatian Pedagogy for Sustainability to energy science courses with a heavy emphasis on community outreach and client engagement.[11] More literature on using the IPP to teach exists in the business literature, applied mostly to data analytics[12]; although some lessons can be garnered from these papers, the context does not include the particular challenges and realities of teaching science. In fact, I found that the majority of the work studying the use of the IPP in STEM disciplines has been at the primary and secondary school levels.[13] Nothing in the literature answers the fundamental question of how to use the IPP to teach university-level STEM courses in an authentic, Ignatian manner.

In my own field of physics, for example, physics education research (PER) has become a well-respected subfield which has thoroughly studied the efficacy of numerous pedagogies in physics courses at all levels, using the scientific method to answer the following question: what evidence would convince me that my students are learning deeply and effectively? However, even though many science and mathematics faculty are familiar with modern teaching methods coming from educational research, most do not see connections with Ignatian Pedagogy or understand how to implement the IPP within their courses. In this article I hope to sketch out a framework that will help STEM faculty adapt and integrate the IPP into their courses and report on the success I have had using the IPP in my own teaching.

**The Ignatian Pedagogical Paradigm**

The Ignatian Pedagogical Paradigm grew out of St. Ignatius's experiences in both participating in and leading the Spiritual Exercises, and the methods and execution are laid out in the Constitutions and in the Ratio Studiorum. Most readers are no doubt familiar with Ignatian Pedagogy, but I give a brief review here to make this article more accessible to colleagues in STEM fields who may have less familiarity with the IPP or for readers at non-Jesuit colleges and universities who are learning about the IPP for the first time. Excellent reviews of the IPP can be found here.[14] At its heart, the IPP is about a three-fold relationship between the learner, the teacher, and the subject matter to be learned, i.e. "truth". The IPP requires a reverence for the subject matter and truth, coupled with a reverence for the student; it asks also that the student gives of himself or herself generously. The teacher in the IPP is not a "sage on the stage", but rather, a guide or a coach. The very essence of the IPP is hence learner-centered; the student, rather than the teacher, is the most important person in the classroom. Learning in the IPP is cyclical and based on a repeated encounter with the material; these encounters deepen students' understanding at each iteration, and these iterations are carefully scaffolded by the instructor to optimize learning. Each encounter with the material is followed by a two-fold response: first, reflection upon the material, and then action. St. Ignatius stressed "itelligo et faciam", or "I understand so that I might act." More on reflection and action will follow later as these are critical components in the IPP.

My work in physics education has exposed me to educational theory and both cognitive and psychological science perspectives on learning. My response to the IPP is always a sense of wonder. St. Ignatius, some 400+ years ago, understand a great deal about learning and the human condition that cognitive science has only been teasing out in modern times. For example, the Kolb Learning Cycle[15], which has been hailed as a scientific theory of learning, stresses a cyclical model of learning in which students learn through experience, reflection upon that experience, abstract conceptualization, and then action based on the previous steps. "Learning is the process whereby knowledge is created through the transformation of experience".[16] This is essentially the IPP with its emphasis on experience, reflection, followed by action. Educational research stressing the need for active-engagement strategies and downplaying the effectiveness of passive lecturing mirrors Ignatius's insistence on learner-centered education. Hence the IPP, though more than 400 years old, nonetheless embodies the "best practices" about teaching and learning of the 21st century.

The IPP in its modern incarnation is presented in most sources divided into the following practical components: 1. Context, 2. Experience, 3. Reflection, 4. Action, and 5. Evaluation. The IPP involves a cyclical process in which students move through stages 2-5 above and then back again.

**Reflection**

I will start with reflection because I believe that reflection is at the heart of the IPP and hence sits at the core of all of my courses. In yet another example of how St. Ignatius was 400+ years ahead of his time, John Dewey, architect of the modern American educational system, also emphasized the importance of reflection; a quote often attributed to Dewey, but which is more of a paraphrase of his educational philosophy, asserts that we do not learn from experience, rather that we learn from reflecting on experience.[17]

Reflection has been written about heavily in the context of Ignatian Pedagogy with some excellent articles exploring its use in undergraduate education.[18] Another excellent resource on reflection is a working paper from the University of Exeter[19] that builds a cognitive science-based model of reflective practice. However, I have also learned a great deal about reflection and how to analyze reflections from the teacher training and education literature. What this literature points out, unsurprisingly, is that students do not begin as the best reflective practitioners. Spalding and Wilson point out, "Reflection is a mysterious concept to many students … few have written - or perhaps even thought - reflectively during their academic careers."[20] In my own courses I find that most physics students react with a "huh?" when asked to reflect in a physics course. Although these students are sometimes more familiar with reflection in the context of humanities or social science courses or through experiences in campus ministry or service, reflection in a STEM course is inherently alien to them, and therefore they require some training and scaffolding. During the first day of class, I spend time discussing reflection and why it is critical to learning. I introduce them to the Ebbinghaus Forgetting Curve, which shows how quickly newly acquired knowledge is forgotten and that deep learning requires frequent refreshing/reflection.[21] I also present a problem-solving strategy for STEM homework called the "Method of Three Passes" which heavily incorporates reflection.[22] And finally, I share with them a short paper from the physics education literature in which an instructor gave the same questions on a midterm and final exam and found that students performed similarly if not worse on the final. That paper concludes, "We suggest that many advanced physics students do not routinely exploit their mistakes in problem solving as a learning opportunity."[23] In other words, most students do not develop the reflective (or metacognitive self-monitoring) skills that are necessary for deep learning. To further sell students on the power of reflection I share researcher David Meltzer's characterization of the most successful physics students that he has encountered in his long career: "highly successful physics students . . . are active learners. They continuously probe their own understanding of a concept …".[24]

To scaffold the reflective process for students, I incorporate written reflections into our campus LMS which are delivered as quizzes. After short, insufficient, and shallow reflections which often had students list their main goal in the course as receiving an "A" grade, I also built in rubrics for all reflective exercises and provide students with examples of what I am looking for. The kinds of reflections I build into my courses are:

1. Beginning of the semester reflection: this reflection has students share details about themselves, their level of preparation, their hopes and worries for the course, and what I can do as an instructor to help them learn. It also asks students to establish 2-3 clear learning goals for themselves for the semester. And finally, it asks students to reflect on what they will do to hold themselves accountable for their own learning and to support meeting their goals.
2. Cognitive Exam Wrappers: Cognitive exam wrappers, or simply exam wrappers, were introduced by Marsha Lovett and colleagues at Carnegie Mellon University in response to what they saw as a failure of students to use exams as learning opportunities.[25] I use a shortened version of Lovett's exam wrapper, due one week after students have received their graded exams back, and ask students to reflect on the following: 1) How they studied for the exam, 2) What mistakes they made on the exam, 3) What they can potentially do differently to prepare more effectively for the next exam, and 4) What I as the instructor can do to help support their learning and help them prepare for the next exam. Exam wrappers give students a concrete and constructive way to think about what went wrong and what they struggled with from the exam. Without this reflection, students rarely give the material from the last exam a second thought, even if they will see it again on a cumulative final.
3. Project Reflections: In courses where I employ projects rather than exams, I replace cognitive exam wrappers with project reflections. I ask: 1) what they learned from the project, 2) what big ideas from the course did the project help them understand better, 3) their role and the role of other group members (teamwork), 4) where they struggled and what they still did not understand, and 5) how could the project be made better or more effective for learning.
4. Weekly Technical Reflections: These are short reflections that ask students to reflect on the week's course material and identify three things: 1) what they learned (and how), 2) what the most important thing they learned was, and 3) What they were struggling with this week. Weekly technical reflections are an excellent resource for instructors to get the pulse of the class and zero in on the topic or topics most students are still struggling with.
5. End of semester reflection: This reflection looks back at the student's initial goals and asks if the student met them or made progress towards meeting them. It asks students to reflect on their experience in the course and whether they developed the relevant course skills, matured as a student, and made progress in taking charge of their own learning.

The scaffolding and the different types of reflections help students grow as reflective practitioners. I analyzed student reflections in several courses using a schema from the educational literature which classifies reflections into five categories, starting with superficial reflections about class time and assignments (what happened) to critical reflection in which students reflect more deeply on how and why they learn and provide constructive criticism to improve the learning environment.[26] I found that students were in fact capable of serious and sophisticated reflection, and that the scaffolding scheme was effective in helping students reflect at higher levels. The reflections therefore contained more than simple narrative data about their experiences and instead gave me new insights into students' attitudes and epistemological states, data that would have been very difficult to capture in other ways.[27]

**Context**

The context element of the IPP reminds us that education is always about the student and never about the instructor. We are called to know our students and to know their hopes, dreams, and aspirations. Gregory Carlson, S.J., once delivered an address to a group of faculty working to create a new science program at a Jesuit university. When asked what such a program would look like if it grew out of Ignatian sensibilities, his first point was that the program should be anchored in "student desires."[28] In other words, instructors need to find out what students want. What are they sacrificing to be there? What are they willing to give up to achieve success? To know these things about our students means knowing them beyond superficial conversations before and after class. I know that as a physicist I am often in awe of my colleagues in the humanities with regards to how easily they draw conversation out of their students and how they often seem to know so much about their students' lives. I do not have that kind of gift, but I have found that I get to know my students in this way because of the reflections I employ. The reflections allow me to get to know my students, their goals, their hopes and fears, and obstacles they may have to learning. In particular, written reflections are even more useful with students who are reticent to talk in class due to social anxiety, autism, or other challenges. Written reflections place every student on a level playing field in terms of access and attention from an instructor. However, this means that I as the instructor must absolutely read and respond to student reflections in a timely manner; students need to feel heard and valued, and the best way to do that is to respond to their reflections with the same generosity.

I think there is another element of context that is particularly important to teaching STEM courses. One of the lessons cognitive science has taught us over the last thirty years is that learning is associative and knowledge is not learned in isolation but is built upon and connected to previous knowledge.[29] In other words, our students are not empty slates but come to us with knowledge and a world-view and a deep-seated intuition about how the world works. Educational research has found that such previous knowledge can get in the way of learning. Unless previous misconceptions are addressed and explicitly confronted, learning can be difficult or impossible. Andrea diSessa, for example, has done brilliant work on what he terms phenological primitives or p-prims; p-prims are these deeply held notions about how the world works.[30] One example would be the following: if a large truck hits a compact car, it ends badly for the compact vehicle. This "victory belongs to the bigger/stronger" p-prim can lead students to incorrectly reason out how Newton's 3rd Law operates in nature, i.e. that in a collision, the forces on both objects are equal but in opposite directions. Other p-prims like "closer means more" (the closer you get to a campfire the hotter it gets) can lead students to incorrectly reason out that summer occurs because the Earth is closer to the Sun in summer than in the winter. Conceptual assessment exams are excellent tools at ferreting out these misconceptions. Once an instructor knows what his or her students actually believe, then the work of education can begin. For example, I have always been amazed by a result from physics education about the efficacy of lecture demonstrations. Thornton and Sokoloff showed that lecture demonstrations, though amusing for students, are absolutely ineffective in promoting learning unless the instructor does one simple thing: have students predict and write down what they thought would happen.[31] The step of making a concrete prediction and then seeing a result that did not concur with the prediction gives students a sense of cognitive dissonance and an opportunity to correct tightly held misconceptions.

**Experience**
Both the IPP and cognitive science tell us that students must engage actively with the material to learn deeply and effectively. Additionally, it has been shown that most people in general require social interaction to learn well. Luckily, educational researchers have developed many active-engagement pedagogies that are suitable for diverse disciplines. Some examples of active-engagement pedagogies that are particularly relevant for STEM instruction include:
1. Flipped-classroom[32] or Just-in-Time Teaching (JITT)[33]: students engage with material at home through study or watching videos/tutorials, and class focuses on a response to student difficulties with the material. These approaches give students the easier to comprehend material to do at home and leave difficulties for class.
2. Peer Instruction/Think-Pair-Share[34]: The instructor poses a conceptual question and polls the class (sometimes using clicker devices or online polling software). If the students answer incorrectly, the instructor has students engage with each other, and then re-polls the class. The instructor can then explain the concept if students are still struggling or move on if they've converged on the correct answer.
3. Interactive Lecture Demonstrations (ILD)[35]: A teacher presents a live demonstration or simulation and engages the class in making predictions and discussing the results of the experiment/simulation.
4. Problem-based Learning (PBL)[36]: Students work in small groups to solve real-world problems that are often vague or open-ended and do not contain all the information needed. A real-world scenario that the students would care about acts as a "hook" to draw students into learning.
5. Inquiry-based Learning (IBL)[37]: This method was developed for use in mathematics courses and consists of students solving problems individually or in groups which they then present to the class. The instructor and peers provide feedback.

6. Modeling Instruction[38]: Students use mathematical software like Mathematica or MATLAB or coding using languages such as Python to solve problems and simulate more realistic, real-world problems.
7. Process Oriented Guided Inquiry Learning (POGIL)[39]: This method, used extensively in chemistry and engineering, has students working in teams on guided tutorials, problems, derivations, etc. Lecture is replaced by these guided activities, which sometimes include laboratory experiments. Workshop Physics[40] is an example of a POGIL-like implementation in physics.

To use Ignatian language, students need to "taste" what they are learning and engage it with all of their faculties, i.e. not just intellectually. Hence teachers should craft many different kinds of educational experiences for students: reading, problem solving, lecture demonstrations, videos, laboratory experiments, etc. For example, I always tell my physics students that physics is a whole-brain activity; something as simple as drawing a picture has been shown to be critical to successfully solving a physics problem[41]. Emotion is also key to learning, as both St. Ignatius and modern cognitive science understand; can we as educators craft scenarios that invoke an emotional response and hence increase learning? Can we pick real-world scenarios that our students actually care about? Can we tie in ethical issues that make the learning of the science more compelling? This is where the invisible scaffolding occurs, and much of the work that a good teacher does happens out of student view.

**Action**

In the IPP, education naturally leads to action, whether that action is changes to a student's attitude and priorities or to changes in a student's life and choices and purpose. In discussions with colleagues in the mathematics and natural sciences, I have found that this is the area most of us struggle with and is the element of the IPP which is most daunting. After all, what kind of action should a course in theoretical physics or a mathematics course on differential equations lead to? I think, though, that letting this area be an impediment to adopting the IPP does a disservice to our students and to ourselves as teachers. I also firmly believe that we influence our students into action far more than we ever know. I have seen pre-medical students fall in love with physics and change their majors. Another student might find that that research is exhilarating and change their entire career trajectory. Students hopefully find that they look at the world differently after a course in the natural sciences. Are these not actions that grew out of our teaching and care for our students? Every time we inspire our students, we set their feet on the path to action.

Taking a science course and adding a service-learning element is a wonderful way to introduce students to practical, external action and to model how students can use what they are learning to become men and women with and for others. Many good examples exist in the sciences, and I will mention a few. One example comes from Creighton University, where courses in the energy science program partnered with community organizations to help research and design sustainable energy installations that would allow those partners to serve their clients more effectively.[42] Another example comes from both Indiana University and Notre Dame, where chemistry students collected and analyzed lead samples as part of a citizen science project to help those in their communities who were affected by lead poisoning, often those at the margins.[43] But not every example of service-learning needs to be so grand. When I was in charge of our physics seminar, which teaches students about science communication, I had students pick simple toys like a yo-yo, spinning top, silly putty, etc. and then partner with a local Jesuit middle school to meet with and explain the science behind the toy to middle school science students. This simple step put our students into a local school which served primarily poor, underprivileged students and got them thinking both about how to communicate science, but also about privilege, poverty, and how they could make an impact.

Not every science course can of course include a service-learning component. More practically speaking, one kind of action that I think is very faithful to Ignatian sensibilities is to be clear with students about what they are learning is "good for." In other words, students should see a very clear connection between theoretical knowledge and what that knowledge can be used to accomplish and/or discover. This might mean a very tight correlation between theoretical calculations and the results of experiments in a particle physics class, always linking what can be theorized to what is actually observed and measured. Or it could mean that the techniques students learn in a course are applied to real-world problems and situations of interest to students. For example, I spent several years writing a weblog for my general physics students in which I related the physics we were learning in class to real-world effects like how the gas pump knows to shut off when your gas tank is full.[44] Homework problems and projects can and should involve real-world connections. Ethical issues that arise as a result of course material should be discussed and not swept under the rug. Closely linking course concepts and deliverables to real-world scenarios shows students how they can contribute to society through that discipline and motivates their learning.

**Evaluation**

The IPP, like modern learning theory, stresses the need for frequent assessment and evaluation to judge whether or not students are learning and to pinpoint what they still struggle with. Evaluation and assessment are where disciplinary-based educational research can be extremely helpful. For example, in physics, researchers have created and nationally normed conceptual assessment exams designed to measure student learning in courses from introductory physics to graduate courses. Tools such as the Force Concept Inventory, created in the early 1990s, measures students' understanding of motion and Newton's laws.[45] In fact, it was a meta-analysis of physics courses evaluated using this instrument by Richard Hake in 1998 that convinced many physicists of the need for active-engagement physics teaching; scores on the FCI demonstrated that students in active-learning classrooms came away from a physics course with a much better understanding of mechanics and motion[46]. In chemistry, the American Chemistry Society creates cumulative conceptual exams and curates national statistics for courses such as general and organic chemistry. But even if such conceptual exams do not exist at the national level for a course, instructors can model home-built assessment instruments from these for their own courses. For example, my home department has used sample questions from the physics graduate record exam (GRE) to as part of the strategy to assess learning in upper division courses where no PER instrument existed.

However, I have also found that my students are themselves rich sources of potential assessment data. As Dennis White points out, *"Asking students to talk about their education is so simple that—whether we are teachers, partners, researchers, or policymakers—we inevitably forget to do it."*[47] Students are not shy about letting instructors know when things have not gone well or when they have felt unprepared for exams or projects. Students can be powerful partners for revising and reforming courses.[48] Moreover, the reflection tools I have described above can be used to capture rich data that goes beyond learning of the course material. This data cannot always be analyzed using quantitative methods, and therefore qualitative analysis can be a powerful tool. This can also be essential in smaller, upper division courses where student numbers are simply too small for quantitative methods to bear fruit. For example, one element of student epistemologies that I care deeply about is how students perceive themselves as developing scientists; do they feel that they are simply novices, or do they increasingly self-identify as a physicist? This belief about themselves has a deep effect on how students learn[49]. Identity cannot be teased out by a Likert scale or other quantitative measure, but it emerges from written reflections, focus groups, one-on-one interviews, etc., all of which require qualitative analysis tools to study. I give an example of qualitative analysis in the following section for a problem-based learning class without lectures.

Besides using student reflections and interviews as a rich source of evaluation and assessment, other potential assessments that will work in a broad variety of disciplines are:
1. CATME[50] or other teaming assessments can gauge how students engage with each other as classmates and teammates and answer questions such as: 1) do students respond with generosity towards each other when facing difficulties? 2) do students readily help each other on homework or projects? 3) do students carry their own weight in group assignments?
2. Attitudinal instruments like the Colorado Learning About Science Survey (CLASS)[51], Maryland Physics Expectation Survey (MPEX)[52], Views About Science Survey (VASS)[53], and Epistemological

Beliefs Assessment for Physics Science (EBAPS)[54] gauge students' feelings and thoughts about both a subject and their own epistemological state.  The CLASS instrument, for example, is available in several versions designed for use in physics, biology, chemistry, and mathematics courses.[55]
3. Other inventories exist that capture students' degree of ownership of learning by measuring aspects of self-directed learning and self-regulated learning.[56]
4. Alongside self-directed and self-regulated learning, a great deal of work has been done on student metacognition, or how students think about their own thinking and learning.[57]

The IPP challenges us to move beyond evaluation of strictly course content and instead worry about who our students are and who they are becoming.

**What Might an Ignatian STEM Course Look like?**

Here I describe three types of different course structures for physics courses at different levels of the curricula, all of which incorporate IPP as their educational heart and soul.  Although these are physics courses, I believe that the structure and methods of these courses can be transferred or applied broadly to other STEM disciplines.  My main goal here is to give examples that readers could see potentially implementing in their own teaching.  One point I hope to convey is that I have found that the IPP works well when stacked with other educational innovations/methods, particularly active-engagement strategies.  In other words, and this is an important point, instructors do not have to choose between the IPP and other best-practice methods from their discipline.  Both can coexist and make a course stronger and more effective together.  I also present some assessment data that comes from nationally normed assessment instruments as evidence that my approach is successful.

1. The Flipped Classroom Approach
My year-long general physics course, primarily taken by freshman physics, math, and pre-engineering students, blends IPP with a flipped classroom approach.  Students complete an assessment during the first week of class along with a math skills assessment.  These assessments help me identify the most common misconceptions and where students will struggle.  The flipped-classroom approach works in the following way.  Students watch pre-lecture videos online and answer checkpoint questions about the material before coming to class.  These pre-lecture videos cover the basic elements of the topic for class that day and are essentially equivalent to reading the textbook ahead of time.  Class time involves my response to their difficulties with the pre-lecture and checkpoint questions, and a deeper dive into the material.  Lecture is broken up by active engagement activities such as think-pair-share questions using clicker devices, group problem solving, interactive demonstrations, and field trips with activities such as riding elevators while standing on scales or recording and analyzing billiard ball collisions on the pool tables in the student center.  Homework is done online using MacMillan's Achieve system[58], which gives students immediate feedback, along with two written problems per week submitted electronically through our LMS.  I have found that this approach works best with freshmen students since they still need significant scaffolding and are not ready for more independence and autonomy.  I use standard written exams (three exams and a cumulative final) consisting of conceptual questions plus workout questions that ask students to solve problems similar (but not the same) to those we have seen in class.  I make a heavy use of reflection: I use both the beginning and end of the semester reflection as well as exam wrappers and weekly technical reflections.  The exam wrappers, for example, draw exams out of static, summative assessment/evaluation back into the learning cycle; exams become both evaluation and learning as students reflect on them and circle back to the content.

At the end of the semester students re-take the initial assessment and I calculate learning gains by topic based on how students did on the post-test vs. the pre-test.  In physics, a standard reporting mechanism is what is called the Hake gain, which simply measures a percentage of what students improved on from pre to post-test (i.e. if they missed ten questions on the pre-test but got five of those right on the post-test, a student's Hake gain would be 50%).  For the FMCE exam, typical Hake gains are around 20% for non-interactive lecture courses, and 51.8% for courses using active-engagement strategies.[59] The long-term average Hake gain for the first semester of general physics at my university is 48.9%, which is comparable to the national average. My IPP-based first semester general physics courses have an average Hake gain of 66.1% for the last four years that I have taught the course. Second semester results, using the Conceptual Survey of Electricity and Magnetism (CSEM) assessment, are similar. The CSEM exam has a national average Hake Gain of approximately 22.4%.[60] The long-term average Hake gain for the second semester of general physics at my university is 33.8%. My IPP-based second semester general physics courses have an average Hake gain of 51.5% for the last four years that I have taught the course. Hake gains on these assessment instruments show that my students are internalizing a concrete understanding of concepts in general physics and that the flipped-classroom + IPP approach is highly effective.

2. Problem-based Learning with Lecture
At the sophomore level I design courses to begin to remove the scaffolding of a typical lecture-based course. One way this can be actualized is to have students work in small groups on projects; during project weeks, work on the projects replaces lecture. I use problem-based learning here, and projects are simply more lengthy, more involved problems. I have found that PBL is extremely compatible with the IPP: students are actively engaged, projects can be tailored to student interest, projects invite reflection, and projects require social interaction and teamwork. In my sophomore level classical mechanics class, we revisit content from the first semester of general physics (mechanics) and add complexity. For example, we re-consider projectile motion, but now add air resistance. More problems have friction included. And finally, the course teaches new mathematical techniques such as Lagrangian and Hamiltonian mechanics that allow students to model complex, real-world systems much more easily than with Newtonian mechanics. In this course I typically use three projects, and these projects replace traditional exams. The first project deals with coordinate transformations and ends with students using rotations to calculate a great circle course from Long Beach, CA to the Pearl Harbor, HI, a very practical application of studying rotations. The second project involves projectile motion with air resistance and requires the use of MATLAB or Python to do numerical calculations. The final project has students model the motion of a real amusement park ride, the Black Widow, a swinging and rotating pendulum-like ride at Kennywood Park in Pittsburgh, PA. In that scenario students take on the role of safety engineers who want to determine the maximum g-forces that a rider will experience while on the Black Widow and thus determine if a larger ride planned in the future is safe. Students submit written reports for all of the projects and reflect heavily at the end of each project both on their contributions, their learning, and on how the team dynamics and group function.

To do assessment for this course I use the Colorado Classical Mechanics/Math Methods Instrument (CCMI)[61]; this instrument is not multiple choice and hence more difficult to score. It involves students doing calculations and solving problems in a variety of areas of classical mechanics. I have only taught this course twice recently, but after finding that students scored dismally on the CCMI and on weekly quizzes, I identified a list of core competencies that I wanted to make sure my students emerged from the course with. The next time I taught the course, I gave students mini competency exams spread throughout the semester; students could re-take the competency exam as many times as needed, but needed to demonstrate mastery of the particular skill to move on in the course and receive credit for the competency. All students in the course successfully demonstrated mastery of all eight competencies, although several students took four or five attempts on the more difficult ones. Questions on the written final exam were designed to fall under the umbrella of the eight competencies as well. I include these details to show how assessment and evaluation were used in my course, but also to point out that when the standard instrument like the CCMI didn't fit my course or students well, I modified the approach using ideas from competency-based learning.[62] Context in the IPP also means acknowledging that structures that worked well for other students may not work for the students I currently have in my class.

3. Problem-based Learning with no Lectures
At the junior/senior level I want students to be ready to take charge of their own learning and hence I remove much of the scaffolding I use in other courses. In my quantum mechanics course, I use a problem-based learning approach coupled with the IPP. Students complete the usual reflections, but instead of using lecture with active engagement strategies, I do away with lecture altogether. Students complete tutorials in class on major topics, complete guided reading assignments in their textbook along with homework, and spend significant time working in small groups on projects. The tutorials walk students step-by-step through a topic and force the students to be active and engaged during class. Each tutorial comes with a list of learning objectives and ends with a self-assessment; in this self-assessment, students evaluate themselves based on the learning objectives, and also complete a short problem to be solved using the knowledge gained from the tutorial. Thus, assessment and evaluation are embedded into almost every class period. Guided reading assignments are chunks of the textbook for students to engage with coupled with my own notes on that section pointing out pitfalls and critical points. And finally, projects form a backbone of the course. Based on student interest, I divide the students into groups and have them tackle real-world applications (problems) of the quantum mechanics we are studying. Some examples include coming to an understanding of how alpha-decay occurs in Uranium (quantum tunneling) and why radioactive atoms exist with such different half-lives. Another group might analyze a dye molecule that emits and absorbs light at a particular wavelength, and by modeling the system figure out why that occurs. I emphasize using theory to model real systems and comparing theory to experimental data. Students write up their results using LaTeX typesetting software in the format of a physics journal for the first and second projects, lead a journal club-like presentation for the third project, and print posters similar to those presented at physics conferences for the fourth project. The deliverables for the course therefore are exactly the kinds of products professionals in their field produce. Each project culminates in a reflection on both their work, their deliverable, and on their teamwork.

In this course I used both qualitative and quantitative tools for the assessment of student learning. Because this course did not employ lecture, I was acutely aware that student success in the course would strongly depend on student attitudes towards learning, particularly student willingness to bear more responsibility for their own learning. A common epistemological framework in physics education is that of Hammer[63] who breaks down one dimension of student beliefs into a continuum between transmission and construction. In other words, do students believe that knowledge should be delivered to them through instruction or instead are they open to the idea that they themselves help construct knowledge? Another dimension of student beliefs in Hammer's work has to do with innate ability vs. effort. That is, do students believe that success in physics was the result of innate natural skill or do they recognize the importance of hard work and effort in learning? Using the qualitative tool of emergent coding as applied to student reflections, focus groups, and one-on-one interviews, I looked for evidence of these attitudes and their effects on engagement in the class. I found student success in the lecture-free course was highly correlated to attitudes of construction rather than transmission of knowledge. One student in particular began the semester with a deeply entrenched transmissionist attitude; they were an excellent student and felt that they learned effectively from lecture, the mode of instruction they were used to. It was gratifying to see that student's belief about lecture and knowledge transmission crack and change during the course of the semester as the student became more open to learning independently.[64]

On the quantitative side, I assessed overall learning of concepts in quantum mechanics through the QMS instrument[65] developed by Zhu and Singh; this instrument, a thirty-one question multiple choice exam, is broken down into distinct topics which allows a better discrimination of what areas of the course students struggled with. It is a difficult assessment; the reported national average post-test score based on usage at seven universities is only 37%, although this figure includes both undergraduates and graduate students who are seeing quantum mechanics for a second time. In four years I taught my undergraduate PBL quantum mechanics course, my students averaged 34.2% on this exam. For comparison, another instructor at my university has been teaching quantum mechanics using a very traditional lecture-based approach; their students' average on the QMS post-test over the last four years of the course is 24.3%. Thus, I feel confident that my use of IPP and PBL to teach this course has resulted in a conceptual understanding of quantum mechanics at least as good as, if not better, than traditional approaches. In addition, PBL quantum mechanics students gain much more: practice working in teams, scientific writing skills, self-directed learning skills as they accept more responsibility for their own learning, and growth in their metacognitive self-monitoring skills through reflection. In humility, I suggest that this is an example of magis in Jesuit education; my students are learning the physics as they would at any university, but they learn much more because of the use of Ignatian pedagogy.

**Jesuit Education: Cura personalis, Magis, and Men and Women with and for others**

I would argue that any course that ignores the broader Jesuit ideals of cura personalis and magis, and does not seek to educate men and women with and for others, is ultimately not faithful to the heart of Ignatian Pedagogy. This is true even if the course follows the letter of the IPP exactly by being active, repetitive, reflective, and thoroughly assessed. Therefore, in all of my courses I work very hard to connect with each student, getting to know them, remembering that I am not just responsible for teaching them physics but am responsible to care for them as human beings. In our current reality that means being sensitive to mental health issues and challenges. I encourage and promote community building, encouraging students to care for one another and to help each other through what most consider a difficult course. This is particularly true for first year students, who are reticent to talk to their fellow classmates and are not used to the idea that homework and studying is best done in groups. Mixing students week-by-week into groups is usually a very effective way to help students get to know each other and build community; by the end of the semester everyone has worked with and gotten know everyone else. Although I teach physics, I do spend a bit of time discussing my courses in the context of Jesuit education and the Catholic Intellectual Tradition. In the Catholic tradition, faith and reason are compatible, which I try to demonstrate as both a scientist and practicing Catholic; many students, like society in general, buy into the popular myth that science and religion are fundamentally incompatible.[66] To combat this worldview, I take a bit of class time once a year to take my students to see the Heritage edition of the St. John's bible that is on campus to view the scientific illuminations, particularly those in the Gospel of John and in Acts of the Apostles. One of these illuminations is the Earth shown from space, and others include spiral galaxies and comets. As Charles L. Curry, S.J. points out, "Jesuit education was founded on the Renaissance and humanist tradition, to which Ignatius and his followers added the greater breadth and former practicality one finds in the Exercises: all creation is good; all learning is good; and everything in the world can help us find and serve God and one another."[67] In recognition of this, I try to show my students that, like St. Ignatius, we can "find God in all things", even and perhaps particularly in physics.

**Conclusion**

In conclusion, Ignatian Pedagogy is not a method of teaching that is limited to use in humanities or social science courses, but instead is a way of teaching that can be extremely effective in the sciences and STEM more broadly, particularly when coupled with active-engagement strategies from the DBER literature. In this way the IPP is, in a sense, an example of magis: it humanizes and grounds STEM, and helps teachers know, educate, and care for students in holistic way. Reflecting on my own teaching, I find that the IPP has helped my students become better learners by opening their eyes to the need for and power of reflection, by scaffolding opportunities for repetition and deeper learning, and by shifting the power dynamics so that my classroom is truly student-centered. I can honestly say that IPP has delivered on the promise to make me a better teacher as well.

My hope is that this article is useful to any STEM faculty that are interested in adopting the IPP but just do not know how or worries that it does not apply to their field, and this in some small way begins to address the paucity of work on the IPP within STEM. I will conclude, as is appropriate in an article discussing the IPP, with a call for action. My hope for the future of Ignatian Pedagogy is this. Within physics, meta-analyses of active-engagement pedagogies using assessment instruments such as the FCI and FMCE have established the validity and effectiveness of active-engagement strategies.[68] Broader studies have established the same in science and engineering.[69] My hope for study of the IPP is that we will soon have a wealth of research on the use of the IPP within STEM fields, and that this that will allow for similar studies and meta-analyses which will demonstrate the power and particular effectiveness of Jesuit education and of the use of the IPP in the sciences. I hope as well that STEM teachers will adopt the IPP not simply because they teach at a Jesuit institution, but because it is effective.

Notes

[1] Diaz, Ernesto. 2025. "Jesuit Education in Mathematics." Unpublished Master's Thesis. https://doi.org/10.33015/dominican.edu/2006.edu.01.

[2] Peter-Hans Kolvenbach, S.J., Address to the Georgetown Board of Directors, Pontifical Gregorian University, May 10, 2007. https://www.educatemagis.org/wp-content/uploads/2015/01/Georgetown-Address-Paper-on-the-4-Cs.pdf

[3] George Sarton, "Preface to Volume 40: An Appeal for the Republication in Book Form of Father Bosmans' Studies on Belgian Mathematics in the Sixteenth and Seventeenth Centuries", *Isis 40*, no. 1 (1949): 3.

[4] Rivka Feldhay, "The Cultural Field of Jesuit Science" in *The Jesuits: Cultures, Sciences, and the Arts, 1540-1773,* ed. John W. O'Malley, S.J., Gauvin Alexander Bailey, Steven J. Harris, and T. Frank Kennedy (University of Toronto Press, 1999), 107-130.

[5] Mordechai Feingold ed*., The New Science and Jesuit Science: Seventeenth Century Perspectives* (Norwell, MA: Kluwer Academic Publishers*,* 2003), vii.

[6] International Commission on the Apostolate of Jesuit Education (ICAJE). *Characteristics of Jesuit Education* (Rome: Jesuit Secretariat for Education, 1986).

[7] ICAJE, *Characteristics,* 2.

[8] International Commission on the Apostolate of Jesuit Education (ICAJE). *Ignatian pedagogy: A practical approach (Rome: Jesuit Secretariat for* Education, 1993).

[9] Inez Ponce de Leon, "Less prescription, more brainstorming: Teaching science and risk communication as a research class," Journal of Pedagogical Sociology and Psychology 6, no. 1 (2024).

[10] Sarah Dysart, "Supporting STEM Faculty in the Transition Online: A Sloan-C Effective Practice." *Distance Education Report* 17, no. 12 (2013): 5–6.

[11] Andrew Baruth, "Ignatian Pedagogy for Sustainability to Support Community-based Projects: Client-focused Sustainable Energy Solutions," *Jesuit Higher Education: A Journal* 8, no. 2 (2019).

[12] Kathleen Garwood Campbell and João Neiva de Figueiredo, "Ignatian Pedagogy in Quantitative Business Courses: Using Experience and Reflection to Enrich Learning in Data Mining." *Journal of Jesuit Business Education* 10, no. 1 (2019): 68–94; Frederick Kaefer, Guillermina Luz Mora, Ravi Nath, and Nicholas J. C. Santos, "Teaching Data for the Greater Good Utilizing the Ignatian Pedagogical Paradigm," *Journal of Jesuit Business Education* 13, no. 1 (2022): 114–28.

[13] Yuliastuti Dwi Lestari and Eko Budi Santoso, "Implementing The Ignatian Pedagogy Paradigm in Mathematics Learning: Annuity Material at The Vocational School Level," *ITM Web Conference 71* (2025): 01007; KC Mok and KLA Chang, "Values Development Through Science Education: Application of Ignatian Pedagogical Paradigm in Science Classrooms", (paper presented at the International Conference on Learning and Teaching, online December 8-10, 2021); Albertus Hartana, "The implementation of Ignatian (reflective) Pedagogical Paradigm strategy for the improvement of students' learning outcomes and motivation in learning natural science for fifth grade students," in *Proceedings of the 2nd International Conference on Education and Training* (2016): 1253-1262; Kevin Quattrin, "Fighting the freeloader effect:

Cooperative learning, attitude, and achievement in a Jesuit secondary math classroom" (Dissertation, University of San Francisco, San Francisco, 2007).

[14] ICAJE, *Ignatian pedagogy: A practical approach*; Johnny Go, S.J. and Rita Atienza. *Learning by Refraction: A Practitioner's Guide to 21st-Century Ignatian Pedagogy* (Manila, the Philippines: Ateneo de Manila University Press, 2019); Zheng (Jessica) Lu and Vicki Rosen, "Practicing Ignatian Pedagogy: A Digital Collection of Resources," *Jesuit Higher Education: A Journal 4*, no. 2, (2015): 135-152.

[15] David A. Kolb. *Experiential learning: Experience as the source of learning and development* (Englewood Cliffs, NJ: Prentice-Hall, 1984).

[16] Kolb. *Experiential Learning*, 38.

[17] Carol Rodgers, "Defining Reflection: Another Look at John Dewey and Reflective Thinking," *Teachers College Record 104,* no. 4 (2002): 842-866.

[18] Susan Mountain and Rebecca Nowacek, "Reflection in Action: A Signature Ignatian Pedagogy for the 21st Century," in *Exploring More Signature Pedagogies: Approaches to Teaching Disciplinary Habits of Mind.* eds. Nancy L Chick, Aeron Haynie, and Regan A R Gurung (Sterling, Va.: Stylus Publishing, 2012), 129-142; Terry Buxton and Nicole Ellison, "Using Ignatian Pedagogy to Guide Reflections While Blogging." *Jesuit Higher Education: A Journal 4*, no. 1 (2015).

[19] Jenny Moon, "PDP Working Paper 4: Reflection in Higher Education Learning" (unpublished manuscript, University of Exeter, Exeter, United Kingdom, 2001). https://www.researchgate.net/publication/255648945_PDP_Working_Paper_4_Reflection_in_Higher_Education_Learning

[20] Elizabeth Spalding and Angene Wilson, "Demystifying Reflection: A Study of Pedagogical Strategies that Encourage Reflective Journal Writing", *Teachers College Record: The Voice of Scholarship in Education 104*, no. 7 (2002).

[21] Herman Ebbinghaus. *Memory: A Contribution to Experimental Psychology,* trans. Henry Ruger and Clara Bussenius (New York: Teachers College, Columbia University, 1913) accessed at https://dn790005.ca.archive.org/0/items/memorycontributi00ebbiuoft/memorycontributi00ebbiuoft.pdf

[22] Robert G. Brown, "The Method of Three Passes", accessed June 20, 2025, http://webhome.phy.duke.edu/~rgb/Class/phy42/phy42/node11.html.

[23] Guangtian Zhu and Chandralekha Singh, "Improving students' understanding of quantum measurement. I. Investigation of difficulties", *Physics Review Special Topics: Physics Education Research 8* (2012).

[24] David Meltzer, "Iowa State University is the New Entrant into Physics Education Research," American Physical Society Forum on Education Newsletter, Summer 1999.

[25] Marsha Lovett, "Using Exam Wrappers to Promote Metacognition" in *Using Reflection and Metacognition to Improve Student Learning* (Stylus Press, Virginia, 2013).

[26] Linda Valli, "Listening to Other Voices: A Description of Teacher Reflection in the United States", *Peabody Journal of Education 72*, no. 1 (1997): 67-88.

[27] Kristina Ward and Gintaras Duda, "The Role of Student Reflection in Project-based Learning Physics Courses," in *Conference Proceedings of the Physics Education Research Conference 2014*, ed., Paula V. Engelhardt, Alice Churukian, and Dyan L. Jones (Minneapolis, MN: American Association of Physics Teachers, 2014).

[28] Greg Carlson, S.J., "What would a Program Look Like if it Grew out of Ignatian Sensibilities?", Creighton University, August 13, 2013.

29 Pooja K. Agarwal and Patrice M. Bain. *Powerful Teaching: Unleash the Science of Learning* (San Francisco, CA: Jossey-Bass, 2019).

30 Andrea A. diSessa, "Toward an epistemology of physics", *Cognition and Instruction 10,* no. 2 (1993): 105-225.

31 David R. Sokoloff and Ronald K. Thornton. *Interactive Lecture Demonstrations: Active Learning in Introductory Physics* (New York City, NY: Wiley, 2006).

32 Michael K. Seery, "Flipped Learning in Higher Education Chemistry: Emerging Trends and Potential Directions," *Chemistry Education Research and Practice 16*, no. 4 (2015): 758-768.

33 Gregor M. Novak, Evelyn T. Patterson, Andrew D. Gavrin, and Wolfgang Christian. *Just-In-Time Teaching: Blending Active Learning with Web Technology* (Upper Saddle River, NJ, Prentice Hall, 1998). Available for download at https://www.physport.org/curricula/jitt/.

34 Eric Mazur. *Peer Instruction: A User's Manual* (Essex, England: Pearson, 2014).

35 Sokoloff and Thornton. *Interactive Lecture Demonstrations,* 4.

36 Barbara J. Dutch, Susan E, Groh, and Deborah E. Allen. *The Power of Problem-based Learning* (Sterling, VA: Stylus Publishing, 2001).

37 The Academy of Inquiry Based Learning, accessed on June 20, 2025, http://www.inquirybasedlearning.org.

38 Aline Abassian, Farshid Safi, Sarah Bush, and Jonathan Bostic, "Five different perspectives on mathematical modeling in mathematics education, "*Investigations in Mathematics Learning 12*, no. 1 (2020): 53-65.

39 Process Oriented Guided Inquiry Learning (POGIL), "What is POGIL?", last modified 2023, https://pogil.org/what-is-pogil.

40 Priscilla Laws, "Millikan lecture 1996: Promoting active learning based on physics education research in introductory physics courses", *American Journal of Physics 65*, no. 1 (1997): 14.

41 Alan Van Heuvelen, "Learning to think like a physicist: A review of research-based instructional strategies", *American Journal of Physics 59* (1991): 891-897.

42 Baruth, "Client-focused Sustainable Energy Solutions".

43 Gabriel M. Filippelli, Jessica Adamic, Deborah Nichols, John Shukle, and Emeline Frix. "Mapping the Urban Lead Exposome: A Detailed Analysis of Soil Metal Concentrations at the Household Scale Using Citizen Science" *International Journal of Environmental Research and Public Health* 15, no. 7 (2018): 1531; Meghanne Tighe, Christopher Knaub, Matthew Sisk, Michelle Ngai, Marya Lieberman, Graham Peaslee, and Heidi Beidinger, "Validation of a screening kit to identify environmental lead hazards," *Environmental Research 181* (2020): 108892.

44 Gintaras Duda and Katherine Garrett , "Blogging in the physics classroom: A research-based approach to shaping students' attitudes towards physics", *American Journal of Physics 76* (2008): 1054.

45 David Hestenes, Malcolm Wells, Gregg Swackhamer, "Force concept inventory" *The Physics Teacher* 30: (1992): 141–166.

46 Richard Hake, "Interactive-Engagement Versus Traditional Methods: A Six-Thousand-Student Survey of Mechanics Test Data for Introductory Physics Courses," *American Journal of Physics 66* (1998): 18809.

[47] Dennis White. Foreword in Kathleen Cushman. *Fires in the mind: What kids can tell us about motivation and mastery* (San Francisco, CA: Jossey-Bass, ix-xi).

[48] Gintaras Duda and Mary Ann Danielson, "Collaborative curricular (re)construction - Tracking faculty and student learning impacts and outcomes five years later", *International Journal for Students as Partners 2*, no. 2 (2018): 39–52.

[49] David Hammer, "Epistemological Beliefs in Introductory Physics", Cognition and Instruction 12, no. 2 (1994) 151-83.

[50] CATME, "Peer Evaluation: The Original Comprehensive Assessment", last modified 2025, https://info.catme.org/features/peer-evaluation/.

[51] Wendy K. Adams, Katherine Perkins, Noah S. Podolefsky, Michael Dubson, Noah D. Finkelstein, and Carl E. Wieman, "New instrument for measuring student beliefs about physics and learning physics: The Colorado Learning Attitudes about Science Survey," *Physical Review Special Topics: Physics Education Research 2*, no. 1 (2006).

[52] Edward F. Redish, Jeffrey Saul, and Richard N. Steinberg, "Student Expectations in Introductory Physics," American Journal of Physics 66, no. 3 (1998): 212-224.

[53] Ibrahim Halloun and David Hestenes, "Interpreting VASS dimensions and profiles for physics students," *Science and Education 7*, no. 6 (1998): 553.

[54] Andrew Elby, "Helping physics students learn how to learn," *American Journal of Physics 69* (2001).

[55] CLASS: "Colorado Learning Attitudes about Science Survey," accessed June 20, 2025, https://www.colorado.edu/sei/class.

[56] Philip C. Abrami, Anne Wade, Vanitha Pillay, Ofra Aslan, Eva M. Bures, Caitlin Bentley, "Encouraging self-regulated learning through electronic portfolios," Canadian Journal of Learning and Technology 34, no. 3 (2008).

[57] Julie Dangremond Stanton, Amanda J. Sebesta, and John Dunlosky, "Fostering Metacognition to Support Student Learning and Performance," *CBE – Life Sciences Education 20*, no. 2 (2021).

[58] MacMillan Achieve System, 2025, https://achieve.macmillanlearning.com/start

[59] Joshua Von Korff, Benjamin Archibeque, K. Alison Gomez, Tyrel Heckendorf, Sarah B. McKagan, Eleanor C. Sayre, Edward W. Schenk, Chase Shepherd, and Lane Sorell, "Secondary analysis of teaching methods in introductory physics: A 50 k-student study", *American Journal of Physics 84,* no. 12 (2016): 969-974.

[60] Philip Eaton, Barrett Frank, Keith Johnson, and Shannon Willoughby, "Comparing exploratory factor models of the Brief Electricity and Magnetism Assessment and the Conceptual Survey of Electricity and Magnetism," *Physical Review Special Topics: Physics Education Research 15* (2019): 020133.

[61] Marcos D. Caballero and Steven J. Pollock, "*Assessing Student Learning in Middle-Division Classical Mechanics/Math Methods*", in *Conference Proceedings of the Physics Education Research Conference 2013*, ed., Paula V. Engelhardt, Alice Churukian, and Dyan L. Jones (Portland, OR: American Association of Physics Teachers, 2013).

[62] Richard Vorhees, "Competency-Based Learning Models: A Necessary Future" *New Directions for Institutional Research 110* (2001): 5 – 13.

[63] D. Hammer, "Epistemological Beliefs," 157.

[64] Gintaras Duda and Kristina Ward, "Student Epistemologies in Project-based Learning Courses" in *Conference Proceedings of the Physics Education Research Conference 2014*, ed., Paula V. Engelhardt, Alice Churukian, and Dyan L. Jones (Minneapolis, MN: American Association of Physics Teachers, 2014).

[65] Guangtian Zhu and Chandralekha Singh, "Surveying students' understanding of quantum mechanics in one spatial dimension," *American Journal of Physics 80* (2012): 252.

[66] Robert McCarthy and John Vitek. *Going, going, gone: The Dynamics of Disaffiliation in Young Catholics* (St. Mary's Press, 2018).

[67] Charles L. Currie, S.J., "Some Distinctive Features of Jesuit Higher Education Today," *Journal of Catholic Higher Education 29*, no. 1 (2010): 113-129

[68] Von Korff et. al, "Secondary analysis", 969-974.

[69] Scott Freeman, "Active Learning Increases Student Performance in Science, Engineering, and Mathematics." *Proceedings of the National Academy of Sciences 111*, no. 23 (2014): 8410–8415.